\documentclass{article}
\usepackage{spconf}

\usepackage{tikz,mathpazo}
\usepackage{amsmath}
\usepackage{amssymb}
\usepackage{graphicx}
\usepackage{caption}
\usepackage{enumerate}
\usepackage{cite}
\usepackage[colorlinks=True]{hyperref}
\usepackage[latin9]{inputenc}
\usepackage{amssymb}
\usepackage{subcaption}
\usepackage[english]{babel}
\usepackage{algorithm}
\usepackage{tikz}
\usepackage[noend]{algpseudocode}
\usetikzlibrary{shapes,arrows}
\usetikzlibrary{automata,shapes}
\usetikzlibrary{arrows.meta}
\usetikzlibrary{arrows}
\usetikzlibrary{arrows,shapes,chains}

\title{
    Sequence-to-sequence Models for Small-Footprint Keyword Spotting
}

\name{\Large Haitong Zhang \qquad Junbo Zhang \qquad Yujun Wang}

\address{\Large Xiaomi Inc., Beijing, China  \\
\{zhanghaitong, zhangjunbo, wangyujun\}@xiaomi.com
}

\begin{document}
\topmargin=0mm

\maketitle

\begin{abstract}

In this paper, we propose a sequence-to-sequence model for keyword spotting (KWS). Compared with other end-to-end architectures for KWS, our model simplifies the pipelines of production-quality KWS system and satisfies the requirement of high accuracy, low-latency, and small-footprint. We also evaluate the performances of different encoder architectures, which include LSTM and GRU. Experiments on the real-world wake-up data show that our approach outperforms the recently proposed attention-based end-to-end model. Specifically speaking, with $\sim$73K parameters, our sequence-to-sequence model achieves $\sim$3.05\% false rejection rate (FRR) at 0.1 false alarm (FA) per hour.

\end{abstract}


\section{INTRODUCTION} \label{introduction}


Keywords Spotting (KWS), recently used as a wake-up trigger in the mobile devices, has become popular. As a wake-up trigger, KWS should satisfy the requirement of small memory and low CPU footprint, with high accuracy.

There are extensive researches on KWS, although most of them do not satisfy the requirements mentioned. For example, some systems \cite{miller2007rapid, mamou2007vocabulary} are used to process the audio database offline. They generate the rich lattices using large vocabulary continuous speech recognition system (LVCSR) and search for the keyword. Another commonly used technique for KWS is the keyword/filler Hidden Markov Model (HMM) \cite{rose1990hidden, rohlicek1989continuous, wilpon1991improvements}. In these models, HMMs are trained separately for the keyword and non-keyword segments. The Viterbi decoding is used to search for the keyword at runtime.

Recently, \cite{chen2014small} proposes a Deep KWS model, whose output is the probability of the sub-word of the keyword. Posterior probability handling is proposed to come up with a confidence score for the detection decision.  Other neural networks, such as convolutional neural network (CNN) \cite{sainath2015convolutional}, are used in the similar model architecture to improve the KWS performance. To further simplify the pipelines of the KWS model, some end-to-end models proposed can predict the probability of the whole keyword directly \cite{bai2016end, arik2017convolutional,shan2018attention}.

In \cite{arik2017convolutional}, the model uses the combination of the convolution layer and recurrent layer to exploit both local temporal/spatial relation and long-term temporal dependencies. But the latency introduced by the window shifting makes it unpractical. \cite{shan2018attention} solves the problem by adopting an attention mechanism. However, there are two potential problems: (1) the sequence-to-one training is different from sequence-to-sequence decoding; (2) the pre-setting of the sliding window of 100 frame is arbitrary. To handle the problems, we propose a sequence-to-sequence KWS model. With the frame-wise alignments, we can train the model in the sequence-to-sequence framework, and simultaneously get rids of the sliding window.

The attention-based model in \cite{shan2018attention} is used as the baseline model and described in Section 2. Our proposed sequence-to-sequence models are detailed in Section 3. The experiment data, setup, and results follow in section 4. Section 5 closes with the conclusion.


\begin{figure*}
\begin{minipage}[t]{0.5\textwidth}
\tikzstyle{startstop} = [rectangle, rounded corners, minimum width = 3cm, minimum height = 0.7cm, text centered, draw = black]
\tikzstyle{startstop1} = [rectangle,  minimum width = .5cm, minimum height = .5cm, text centered, draw = black]
\tikzstyle{startstopw} = [rectangle,  minimum width = .7cm, minimum height = .6cm, text centered, draw = white]
\tikzstyle{startstop2} = [rectangle, rounded corners, minimum width = 5.5cm, minimum height = .7cm, text centered, draw = black]
\tikzstyle{startstop2w} = [rectangle, rounded corners, minimum width = 5.5cm, minimum height = .7cm, text centered, draw = white]
\tikzstyle{startstop2s} = [rectangle, rounded corners, minimum width = 3.5cm, minimum height = .6cm,align=left, draw = white]
\tikzstyle{startstop2sb} = [rectangle, rounded corners, minimum width = 3.5cm, minimum height = .6cm,align=left, draw = black]
\tikzstyle{startstop2p} = [rectangle, rounded corners, minimum width = .5cm, minimum height = .63cm, text centered, draw = white]
\tikzstyle{arrow} = [thick, -, >= stealth]
\tikzstyle{arrow2} = [thick, ->, >= stealth]
\centering
\begin{tikzpicture}[->,>=stealth',shorten >=1pt,auto,node distance=1.0cm,
                    thick,main node/.style={rectangle,draw,minimum width=.5cm, minimum height=.5cm}]
  \node[main node] (1) {$x_1$};
  \node[main node] (2) [right of=1] {$x_2 $};
  \node[main node] (3) [right of=2] {$x_3$};
  \node[startstopw] (4) [right of=3] {$...$};
  \node[main node] (5) [right of=4] {$x_T$};

  \node[startstop2p] (1p) [above of = 1]{};
  \node[startstop2p] (2p) [above of = 2]{};
  \node[startstop2p] (3p) [above of = 3]{};
  \node[startstop2p] (4p) [above of = 4]{};
  \node[startstop2p] (5p) [above of = 5]{};
  \node[startstop2] (6) [above of=3] {Encoder};

  \node[main node] (9) [above of=6] {$h_3$};
  \node[main node] (8) [left of= 9] {$h_2$};
  \node[main node] (7) [left of =8]{$h_1$};
  \node[startstopw] (10) [right of=9] {$...$};
  \node[main node] (11) [right of=10] {$h_T$};

  \node[startstop2p] (7p) [above of = 7]{};
  \node[startstop2p] (8p) [above of = 8]{};
  \node[startstop2p] (9p) [above of = 9]{};
  \node[startstop2p] (10p) [above of = 10]{};
  \node[startstop2p] (11p) [above of = 11]{};

  \node[startstop2] (12) [above of = 9] {Attention Mechanism};
\node[main node] (13) [above of=12] {$a_2$};
  \node[main node] (14) [left of= 13] {$ a_2 $};
  \node[main node] (15) [left of =14]{$a_1$};
  \node[startstopw] (16) [right of=13] {$...$};
  \node[main node] (17) [right of=16] {$a_T$};

  \node[startstop2s] (18) [above of = 13] {$C= \sum_{t=1}^T a_t h_t $};


  \node[startstopw] (20) [above of = 18] {$p(y)$};

  \draw[->] (1) -- (1p);
  \draw[->] (2) -- (2p);
  \draw[->] (3) -- (3p);
  \draw[->] (5) -- (5p);
  \draw[->] (1p) -- (7);
  \draw[->] (2p) -- (8);
  \draw[->] (3p) -- (9);
  \draw[->] (5p) -- (11);
  \draw[->] (7) -- (7p);
  \draw[->] (8) -- (8p);
  \draw[->] (9) -- (9p);
  \draw[->] (11) -- (11p);
    \draw[->] (7p) -- (15);
  \draw[->] (8p) -- (14);
  \draw[->] (9p) -- (13);
  \draw[->] (11p) -- (17);
     \draw[->] (15) -- (18);
  \draw[->] (14) -- (18);
  \draw[->] (13) -- (18);
  \draw[->] (17) -- (18);
  \draw[->] (18) -- (20);
\end{tikzpicture}
\caption{The baseline model, which is proposed in \cite{shan2018attention}.}
\label{baseline}
\end{minipage}
\begin{minipage}[t]{0.5\textwidth}
\tikzstyle{startstop} = [rectangle, rounded corners, minimum width = 3cm, minimum height = 0.7cm, text centered, draw = black]
\tikzstyle{startstop1} = [rectangle,  minimum width = .5cm, minimum height = .5cm, text centered, draw = black]
\tikzstyle{startstopw} = [rectangle,  minimum width = .7cm, minimum height = 1.0cm, text centered, draw = white]
\tikzstyle{startstop2} = [rectangle, rounded corners, minimum width = 7.0cm, minimum height = 1cm, text centered, draw = black]
\tikzstyle{startstop2w} = [rectangle, rounded corners, minimum width = 7.5cm, minimum height = 1cm, text centered, draw = white]
\tikzstyle{startstop2wl} = [rectangle, rounded corners, minimum width = 5.5cm, minimum height = 1cm,align=left, draw = white]
\tikzstyle{startstop2p} = [rectangle, rounded corners, minimum width = .7cm, minimum height = .93cm, text centered, draw = white]
\tikzstyle{startstophide} = [rectangle, rounded corners, minimum width = .1cm, minimum height = .1cm, text centered, draw = white]
\tikzstyle{arrow} = [thick, -, >= stealth]
\tikzstyle{arrow2} = [thick, ->, >= stealth]
\centering
\begin{tikzpicture}[->,>=stealth',shorten >=1pt,auto,node distance=1.5cm,
                    thick,main node/.style={rectangle,draw,minimum width=.7cm, minimum height=.7cm}]


  \node[main node] (1) {$x_1$};
  \node[main node] (2) [right of=1] {$x_2 $};
  \node[main node] (3) [right of=2] {$x_3$};
  \node[startstopw] (4) [right of=3] {$...$};
  \node[main node] (5) [right of=4] {$x_T$};

  \node[startstop2p] (1p) [above of = 1]{};
  \node[startstop2p] (2p) [above of = 2]{};
  \node[startstop2p] (3p) [above of = 3]{};
  \node[startstop2p] (4p) [above of = 4]{};
  \node[startstop2p] (5p) [above of = 5]{};

  \node[startstop2] (6) [above of=3] {Encoder};

  \node[main node] (9) [above of=6] {$h_3$};
  \node[main node] (8) [left of= 9] {$h_2$};
  \node[main node] (7) [left of =8]{$h_1$};
  \node[startstopw] (10) [right of=9] {$...$};
  \node[main node] (11) [right of=10] {$h_T$};

  \node[startstopw] (18) [above of=9] {$y_3$};
  \node[startstopw] (19) [left of= 18] {$ y_2 $};
  \node[startstopw] (20) [left of =19]{$y_1$};
  \node[startstopw] (21) [right of=18] {$...$};
  \node[startstopw] (22) [right of=21] {$y_T$};

  \draw[->] (1) -- (1p);
  \draw[->] (2) -- (2p);
  \draw[->] (3) -- (3p);
  \draw[->] (5) -- (5p);
  \draw[->] (1p) -- (7);
  \draw[->] (2p) -- (8);
  \draw[->] (3p) -- (9);
  \draw[->] (5p) -- (11);
  \draw[->] (7) -- (20);
  \draw[->] (8) -- (19);
  \draw[->] (9) -- (18);
  \draw[->] (11) -- (22);

\end{tikzpicture}
\caption{The proposed sequence-to-sequence model.}
\label{propose}
\end{minipage}
\end{figure*}

\section{The baseline Model} \label{base}

The baseline model, as shown in Fig. \ref{baseline}, mainly consists of two parts: the encoder and the attention mechanism. 

The encoder learns the higher representation $h=\{h_1, h_2,...,h_T\}$ from the input features $x=\{x_1,x_2,...,x_T\}$. $T=189$ is applied when training. Only LSTM \cite{hochreiter1997long}and GRU \cite{chung2014empirical} are used in our experiments for pair comparison. An attention mechanism \cite{chowdhury2017attention} is applied to come up with an attention weight vector $a=\{a_1,a_2,a_3,...,a_T\}$. Then $C$ is the feature representation for the whole sequential input, which is computed as the weighted sum of $h=\{h_1, h_2,...,h_T\}$. Finally the probability of the keyword $P(y)$ is predicted by a linear transformation and softmax function.


At runtime, the attention mechanism is applied to only 100 frames of input, but only one frame is fed into the network at each time-step since the others are computed already.


\section{The Proposed Model}

The proposed model is illustrated as Fig.\ref{propose}, which mainly includes the sequence-to-sequence training and the decode smoothing.

\subsection{Sequence-to-sequence Training}
The proposed model adopts the sequence-to-sequence training \cite{prabhavalkar2017comparison, chiu2018state}, where the inputs are the features, and the outputs are the one-hot labels which indicate whether the current frame (together with the previous frames) includes the keyword or not.

\begin{figure}[h!]
\centering
\includegraphics[width=0.4\textwidth, height=0.13\textheight]{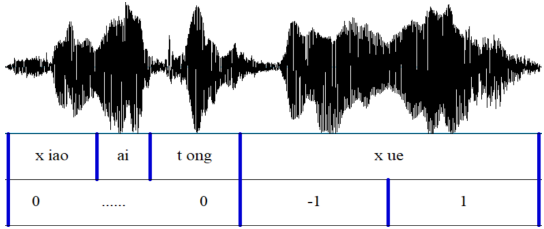}
\caption{The example labeling of the keyword, where the first tire are the alignments and the second the labels.}
\label{labelling}
\end{figure}

An example of labeling the keyword is provided as Fig. \ref{labelling}. Tier one in Fig. \ref{labelling} shows the phone-state alignments generated by the TDNN-LSTM model, which is trained using $\sim$3000 hours of speech. Then the alignments are converted into the one-hot labels. As a result, the frames, which do not include the entire keyword, are labeled as 0. Otherwise, they are 1. The frames are labeled as -1 if they contain three and a half characters. Since these frames are ambiguous, labeling them as -1 and attaching zero weight to them can avoid the potential impact of labeling mistakes.

\subsection{Decoding}
When testing, the model takes the features for a single frame as input, and directly outputs the probability of detecting the keyword $y_t$ . While it is fine to rely on the probability for a single frame, we adopt a smoothing method to come up with a more reliable probability $\hat{y}$, namely the average probability of probabilities of $n$ consecutive frames:

\begin{equation} \label{equation}
\centering \hat{y_t} = \frac{\sum_{i = t-n+1}^t {y_i}}{n}
\end{equation}

\section{Experiment}
\subsection{Dataset}
The keyword in our experiments is a four-Chinese-character term ("xiao-ai-tong-xue"). The training data consists of $\sim$188.9k examples of the keyword ($\sim$99.8h) and $\sim$1007.4k negative examples ($\sim$1581.8h). The development data includes $\sim$9.9K positive examples and $\sim$53.0K negative examples. The testing data includes $\sim$28.8k keyword examples ($\sim$15.2h) and $\sim$32.8k non-keyword ($\sim$37h). The data is all collected from MI AI Speaker \footnote{https://www.mi.com/aispeaker/}.

\subsection{Experiment setup}

40-dimensional filterbank features are computed from each audio frame, with 25ms window size and 10ms frameshift. Then the filterbank features are converted into the per-channel energy normalization (PCEN) \cite{wang2017trainable} Mel-spectrograms.

The cross entropy is used as the loss function in the experiments. While training, all the weight matrices are initialized with the normalized initialization, and the bias vectors are initialized to 0 \cite{glorot2010understanding}. Adam optimizer \cite{kingma2014adam} is used to update the training parameters, with the initialize learning rate of 1e-3. The batch size is 64. Gradient norm clipping to 1 is applied, and L2 weight decay is 1e-5.

\begin{table}
\caption{Performance comparison between the baseline models and the proposed seq-to-seq models, False Reject Rate (FRR) is at 0.1 false alarm (FA) per hour.}
\centering
\begin{tabular}{c c c}
\hline
Model & FRR(\%) & Params(K)\\ [1ex] 
\hline\hline
Baseline GRU &4.47 & 77.5 \\
Baseline LSTM&11.86 & 103  \\
\hline
Seq-to-seq GRU & 3.05 &  73.3   \\
Seq-to-seq LSTM & 6.08 &  86.8   \\ [1ex]
\hline
\end{tabular}
\label{table:comparison}
\end{table}

\begin{figure}
  \includegraphics[width=0.48\textwidth, height=0.30\textheight]{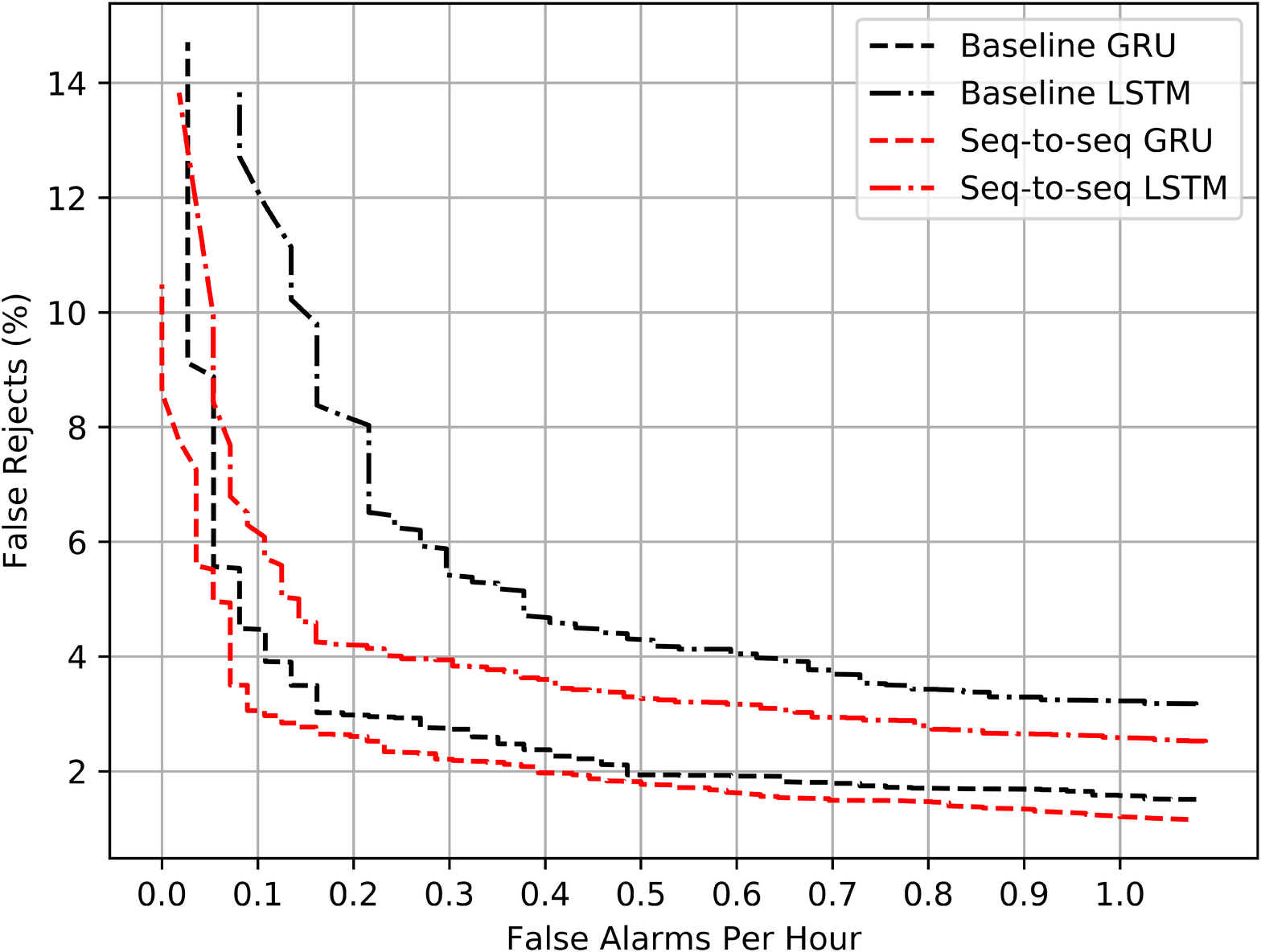}
  \caption{The ROC of the baseline models and the proposed sequence-to-sequence model with the smoothing frame $n$=12. }
  \label{result0}
\end{figure}

\begin{figure}
  \includegraphics[width=0.5\textwidth, height=0.35\textheight]{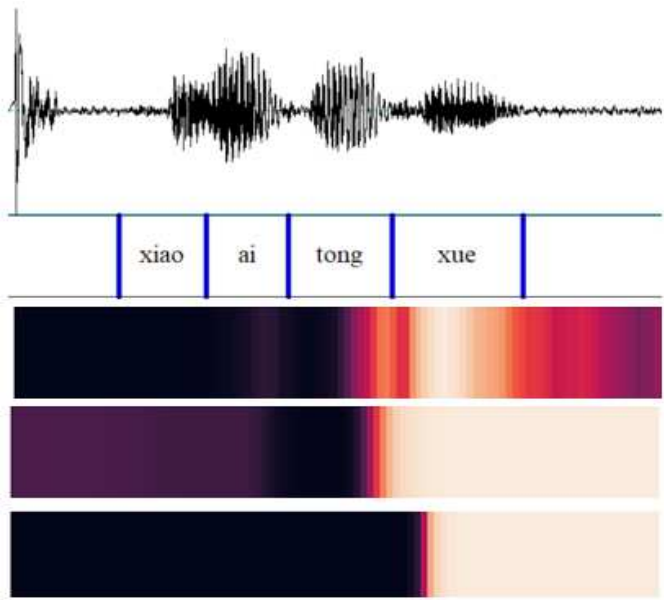}
  \caption{The representative example for the keyword with four tiers of annotation. The first one is the alignment for the keyword;the second is the heatmap for the attention weights learned in Baseline GRU; the third and forth are the heatmaps for the output probabilities given by Baseline GRU and Seq-to-seq GRU, respectively. Larger values are illustrated lighter.}
  \label{heatmat}
\end{figure}

\subsection{Baseline vs Sequence-to-sequence }
The experimental results are reported in the form of Receiver Operating Characteristic (ROC) curve, which is created by plotting the false reject rate (FRR) against false alarm (FA) number per hour at various thresholds. Lower curve represents the better result.

Fig. \ref{result0} illustrates the performance of the baseline models and the proposed models. In this experiment, the encoder is the 1-128 RNN layer, which is found to be the best architecture in \cite{shan2018attention}. It is clearly shown that our proposed model outperforms the baseline models in both LSTM and GRU architectures. The Seq-to-seq GRU achieves $\sim$3\% FFR at 0.1 FA per hour, with an $\sim$20\% improvement over Baseline GRU. The similar situation is observed in the LSTM architecture.

The second tire in Fig. \ref{heatmat} shows that the attention weights concentrate around the last character of the keyword. This distribution indicates that the attention mechanism is strengthening the role of RNN in learning the long-term dependency, rather than focusing on the keyword "with high resolution" \cite{shan2018attention}.


The last heat-mat in Fig. \ref{heatmat} illustrates that sequence-to-sequence model is modeling the human auditory attention.  As people wake up when the entire keyword is perceived, the probability gets large at present of the whole keyword. Although the second heat-mat in Fig.\ref{heatmat} shows a similar picture, the probabilities at the beginning are unreasonably larger than those for the third character, and the wake-up is triggered before the last character is perceived. These impacts can be attributed to two potential problems (Section \ref{introduction}). Instead of using human intervention to set the sliding window for the attention mechanism, our proposed models learn the information implicitly in the sequence-to-sequence architecture.

\begin{table}
\caption{Performance comparison between different encoder architectures, False Reject Rate (FRR) is at 0.1 false alarm (FA) per hour.}
\centering
\begin{tabular}{c c c c c}
\hline
Type &Layer &Unit &FRR(\%) & Params(K)\\ [1ex] 
\hline\hline
LSTM&1& 64 & 7.71 & 27.0\\
LSTM&2& 64 & 7.16 & 60.0 \\
LSTM&3& 64 & 6.55 & 93.1 \\
LSTM&1& 128  & 6.08 & 86.8 \\
\hline
GRU & 1 &  64& 7.79 &  24.5 \\
GRU & 2 &  64& 6.40 &  49.2\\
GRU & 3 &  64& 4.04 & 74.0 \\
GRU & 2 &  128& 3.05 &  73.3 \\[1ex]
\hline
\end{tabular}
\label{table:comparison_encoder}
\end{table}

\begin{figure}[ht!]
  \includegraphics[width=0.5\textwidth]{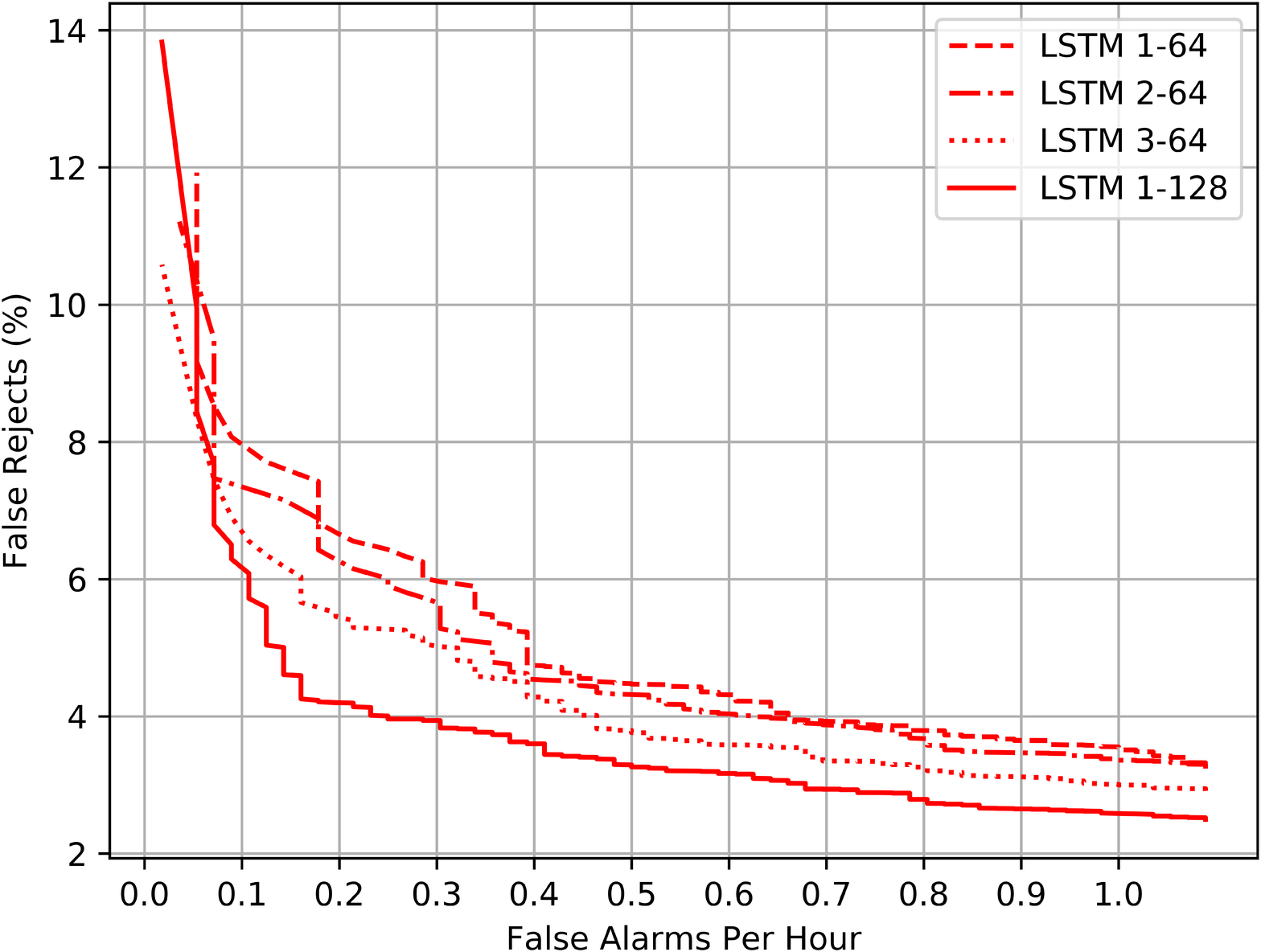}
  \caption{The ROC of the seq-to-seq model with LSTM layer, with the smoothing frame $n=12$. }
  \label{resultlstm}
\end{figure}

\subsection{Impact of encoder}

\begin{figure}
  \includegraphics[width=0.5\textwidth]{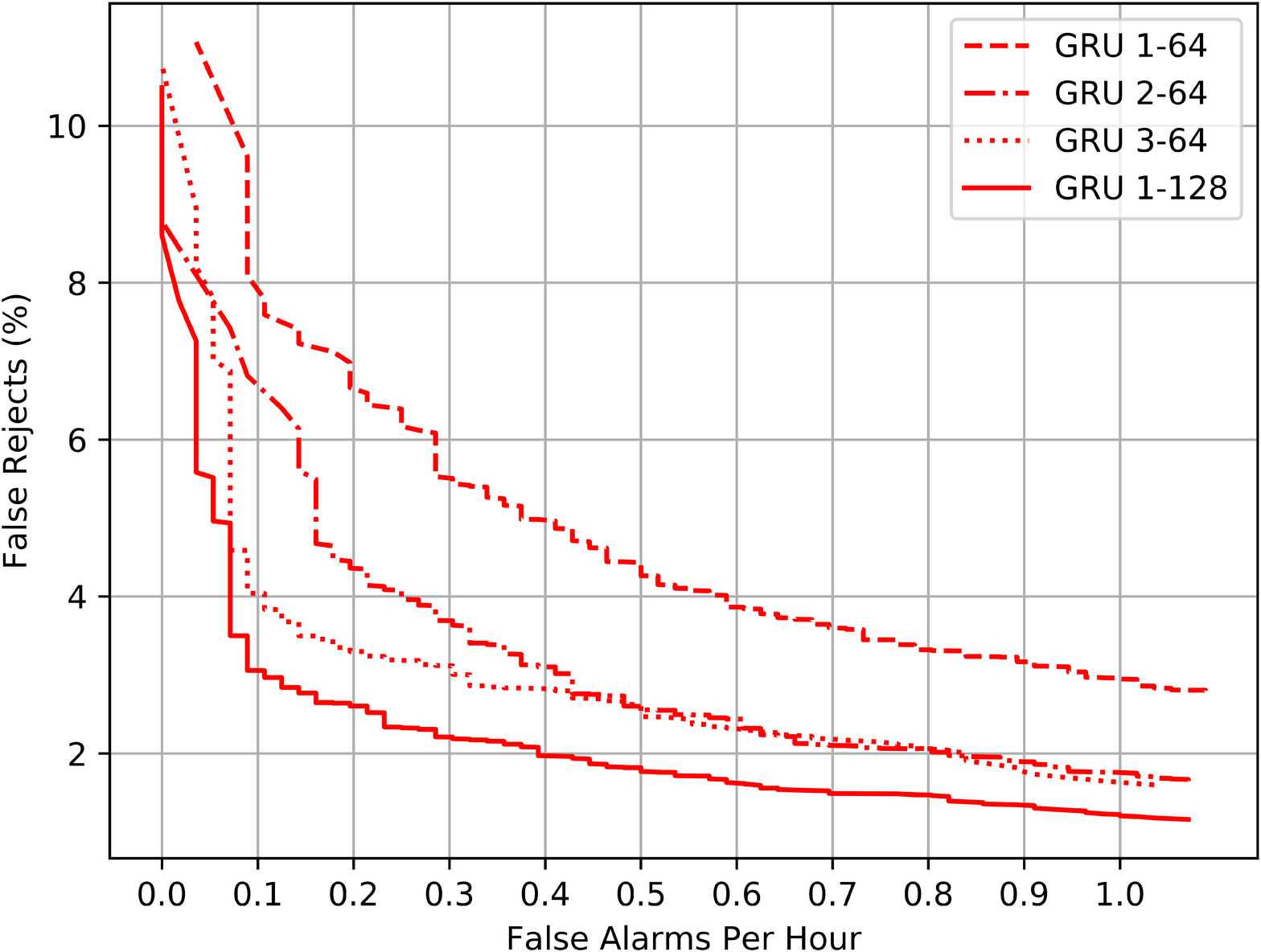}
  \caption{The ROC of the seq-to-seq models with GRU layer, with the smoothing frame $n=12$. }
  \label{resultgru}
\end{figure}

We also explore the impact of the encoder on the model performance. As shown in  Fig. \ref{resultlstm} and Fig. \ref{resultgru}, and Table \ref{table:comparison_encoder}, the models with more parameters tend to perform better than those with fewer parameters. The best models are LSTM 1-128 and GRU 1-128, respectively. As shown in Fig. \ref{resultlstm} and Fig. \ref{resultgru}, the 1-128 models outperform all the models with only 64 units by a large margin, which indicates that getting the network wider results in a better performance than getting it deeper in our experiment.

\subsection{Impact of smoothing frame}
The results of different settings of the smoothing frame $n$ are illustrated in Table \ref{table:comparison_smooth}. Compared with no smoothing, the application of smoothing frame $n=12$ can gain an absolute $\sim$0.15\% and $\sim$0.06\%, respectively in LSTM 1-128 and GRU 1-128. Although the performance difference is minor, we insist that the smoothing strategy is reasonable and pragmatic. It is reasonable because in our sequence-to-sequence model, the detection of the keyword must be kept triggered for several frames once triggered. It is pragmatic since it is computationally cheap to take an average operation.

\begin{table}
\caption{Performance differences due to the smoothing frame, False Reject Rate (FRR) is at 0.1 false alarm (FA) per hour.}
\centering
\begin{tabular}{c c c c}
\hline
Model &Smooth Frame &FRR(\%) \\ [1ex] 
\hline\hline
LSTM 1-128 &1& 6.23  \\
LSTM 1-128 &2& 6.23  \\
LSTM 1-128&5& 6.25  \\
LSTM 1-128 &12& 6.08    \\
\hline
GRU 1-128 & 1 &  3.11    \\
GRU 1-128 & 2 &  3.11    \\
GRU 1-128 & 5 &  3.12 \\
GRU 1-128 & 12 &  3.05\\[1ex]
\hline
\end{tabular}
\label{table:comparison_smooth}
\end{table}

\section{conclusion}

To conclude, the sequence-to-sequence model is more flexible than the attention-based one, because no sliding window is used and the training and decoding strategies are the same. As a result, the proposed sequence-to-sequence model outperforms the other in our real-world data, even with less model parameters. In addition, a computationally-cheap probability smoothing method can improve the performance's robustness.


\begin{thebibliography}{10}

\bibitem{miller2007rapid}
David~RH Miller, Michael Kleber, Chia-Lin Kao, Owen Kimball, Thomas Colthurst,
  Stephen~A Lowe, Richard~M Schwartz, and Herbert Gish.
\newblock Rapid and accurate spoken term detection.
\newblock In {\em Eighth Annual Conference of the International Speech
  Communication Association}, 2007.

\bibitem{mamou2007vocabulary}
Jonathan Mamou, Bhuvana Ramabhadran, and Olivier Siohan.
\newblock Vocabulary independent spoken term detection.
\newblock In {\em Proceedings of the 30th annual international ACM SIGIR
  conference on Research and development in information retrieval}, pages
  615--622. ACM, 2007.

\bibitem{rose1990hidden}
Richard~C Rose and Douglas~B Paul.
\newblock A hidden markov model based keyword recognition system.
\newblock In {\em Acoustics, Speech, and Signal Processing, 1990. ICASSP-90.,
  1990 International Conference on}, pages 129--132. IEEE, 1990.

\bibitem{rohlicek1989continuous}
J~Robin Rohlicek, William Russell, Salim Roukos, and Herbert Gish.
\newblock Continuous hidden markov modeling for speaker-independent word
  spotting.
\newblock In {\em Acoustics, Speech, and Signal Processing, 1989. ICASSP-89.,
  1989 International Conference on}, pages 627--630. IEEE, 1989.

\bibitem{wilpon1991improvements}
JG~Wilpon, LG~Miller, and P~Modi.
\newblock Improvements and applications for key word recognition using hidden
  markov modeling techniques.
\newblock In {\em Acoustics, Speech, and Signal Processing, 1991. ICASSP-91.,
  1991 International Conference on}, pages 309--312. IEEE, 1991.

\bibitem{chen2014small}
Guoguo Chen, Carolina Parada, and Georg Heigold.
\newblock Small-footprint keyword spotting using deep neural networks.
\newblock In {\em Acoustics, Speech and Signal Processing (ICASSP), 2014 IEEE
  International Conference on}, pages 4087--4091. IEEE, 2014.

\bibitem{sainath2015convolutional}
Tara~N Sainath and Carolina Parada.
\newblock Convolutional neural networks for small-footprint keyword spotting.
\newblock In {\em Sixteenth Annual Conference of the International Speech
  Communication Association}, 2015.

\bibitem{bai2016end}
Ye~Bai, Jiangyan Yi, Hao Ni, Zhengqi Wen, Bin Liu, Ya~Li, and Jianhua Tao.
\newblock End-to-end keywords spotting based on connectionist temporal
  classification for mandarin.
\newblock In {\em Chinese Spoken Language Processing (ISCSLP), 2016 10th
  International Symposium on}, pages 1--5. IEEE, 2016.

\bibitem{arik2017convolutional}
Sercan~O Arik, Markus Kliegl, Rewon Child, Joel Hestness, Andrew Gibiansky,
  Chris Fougner, Ryan Prenger, and Adam Coates.
\newblock Convolutional recurrent neural networks for small-footprint keyword
  spotting.
\newblock {\em arXiv preprint arXiv:1703.05390}, 2017.

\bibitem{shan2018attention}
Changhao Shan, Junbo Zhang, Yujun Wang, and Lei Xie.
\newblock Attention-based end-to-end models for small-footprint keyword
  spotting.
\newblock {\em arXiv preprint arXiv:1803.10916}, 2018.

\bibitem{hochreiter1997long}
Sepp Hochreiter and J{\"u}rgen Schmidhuber.
\newblock Long short-term memory.
\newblock {\em Neural computation}, 9(8):1735--1780, 1997.

\bibitem{chung2014empirical}
Junyoung Chung, Caglar Gulcehre, KyungHyun Cho, and Yoshua Bengio.
\newblock Empirical evaluation of gated recurrent neural networks on sequence
  modeling.
\newblock {\em arXiv preprint arXiv:1412.3555}, 2014.

\bibitem{chowdhury2017attention}
FA~Chowdhury, Quan Wang, Ignacio~Lopez Moreno, and Li~Wan.
\newblock Attention-based models for text-dependent speaker verification.
\newblock {\em arXiv preprint arXiv:1710.10470}, 2017.

\bibitem{prabhavalkar2017comparison}
Rohit Prabhavalkar, Kanishka Rao, Tara~N Sainath, Bo~Li, Leif Johnson, and
  Navdeep Jaitly.
\newblock A comparison of sequence-to-sequence models for speech recognition.
\newblock In {\em Proc. Interspeech}, pages 939--943, 2017.

\bibitem{chiu2018state}
Chung-Cheng Chiu, Tara~N Sainath, Yonghui Wu, Rohit Prabhavalkar, Patrick
  Nguyen, Zhifeng Chen, Anjuli Kannan, Ron~J Weiss, Kanishka Rao, Ekaterina
  Gonina, et~al.
\newblock State-of-the-art speech recognition with sequence-to-sequence models.
\newblock In {\em 2018 IEEE International Conference on Acoustics, Speech and
  Signal Processing (ICASSP)}, pages 4774--4778. IEEE, 2018.

\bibitem{wang2017trainable}
Yuxuan Wang, Pascal Getreuer, Thad Hughes, Richard~F Lyon, and Rif~A Saurous.
\newblock Trainable frontend for robust and far-field keyword spotting.
\newblock In {\em 2017 IEEE International Conference on Acoustics, Speech and
  Signal Processing (ICASSP)}, pages 5670--5674. IEEE, 2017.

\bibitem{glorot2010understanding}
Xavier Glorot and Yoshua Bengio.
\newblock Understanding the difficulty of training deep feedforward neural
  networks.
\newblock In {\em Proceedings of the thirteenth international conference on
  artificial intelligence and statistics}, pages 249--256, 2010.

\bibitem{kingma2014adam}
Diederik~P Kingma and Jimmy Ba.
\newblock Adam: A method for stochastic optimization.
\newblock {\em arXiv preprint arXiv:1412.6980}, 2014.

\end{thebibliography}


\end{document}